\documentclass[preprint,aps, showpacs]{revtex4}
\usepackage{graphicx}
\begin{document}
\draft
\title{Partial Wave Analysis of Scattering with Nonlocal Aharonov-Bohm Effect}
\author{De-Hone Lin\thanks{%
e-mail: dhlin@mail.nctu.edu.tw}}
\address{Department of Applied Mathematics\\
National Chiao Tung University, Hsinchu, 30043, Taiwan}
\date{\today }
\begin{abstract}
Partial wave theory of a two dimensional scattering problem for an arbitray
short range potential and a nonlocal Aharonov-Bohm magnetic flux is
established. The scattering process of a ``hard disk'' like potential and
the magnetic flux is examined. Since the nonlocal influence of magnetic flux
on the charged particles is universal, the nonlocal effect in hard disk case
is expected to appear in quite general potential system and will be useful
in understanding some phenomena in mesoscopic phyiscs.
\end{abstract}
\maketitle
\tolerance=10000
In this paper we study the scattering amplitude and the cross section of a
charged particle moving in a short range potential with scattering center
located at the origin, and the Aharonov-Bohm ($AB$) magnetic flux along
z-axis \cite{1}. The nonintegrable phase factor (NPF) \cite{2,3,4} is used
to couple the magnetic flux with the particle angular momentum such that the
partial wave method can be conveniently developed \cite{5,6,7,8,9}. As a
realization of the method, a charged particle scattered by a ``hard disk''
like potential plus the magnetic flux is discussed in detail. Several
interesting nonlocal effects of the magnetic flux in the hard disk model are
concluded as follows: (1) In the long wave length limit ( equivalently,
short range potential) the total cross section is drastically suppressed at
quantized magnetic flux $\Phi =(2n+1)\Phi _{0}/2,$ where $n=0,1,2,\cdots $,
and $\Phi _{0}$ is the fundamental magnetic flux quantum $hc/e$. The global
influence of the magnetic flux on the cross section is manifested with $\Phi
_{0}$ periodicity. On the other hand, the cross section approaches the
flux-free case in the short wave length limit, i.e. the quantum interference
feature of the nonlocal effect gradually disappears, and the cross section
approaches the classical limit. (2) If the hard disk is used to simulate the
boson (fermion) moving in two dimensional space, the scattering process of
identical particles carrying the magnetic flux shows that the total cross
section is suppressed at quantized magnetic flux $\Phi =(2n+1)\Phi _{0}$ for
bosons ($\Phi =2n\Phi _{0}$ for fermions) and exhibits the global structure
with $2\Phi _{0}$ periodicity. Since the nonlocal influence of the magnetic
flux on the charged particle are universal, the influences should be general
in similar systems, and may be useful in understanding some transport
phenomena in mesoscopic physics and account for the quantum Hall effect \cite
{6,10}.

We consider a two dimensional model. The fixed-energy Green's function $%
G^{0}({\bf x,x}^{\prime };E)$ for a charged particle with mass $\mu $
propagating from ${\bf x}^{\prime }$ to ${\bf x}$ satisfies the Schr\"{o}%
dinger equation 
\begin{equation}
\left\{ E-\left[ -\frac{\hbar ^{2}\nabla ^{2}}{2\mu }+V({\bf x})\right]
\right\} G^{0}({\bf x,x}^{\prime };E)=\delta ({\bf x-x}^{\prime }),
\label{01}
\end{equation}
where $V({\bf x})$ is the scalar potential and ${\bf x}$ is the two
dimensional coordinate vector. In the cylindrically symmetric system, the
Green's function can be decomposed as 
\begin{equation}
G^{0}({\bf x,x}^{\prime };E)=\sum_{m=-\infty }^{\infty }G_{m}^{0}(\rho ,\rho
^{\prime };E)\frac{e^{im(\varphi -\varphi ^{\prime })}}{2\pi }  \label{001}
\end{equation}
with $(\rho ,\varphi )$ being the polar coordinates in two dimensional space
and $G_{m}^{0}(\rho ,\rho ^{\prime };E)$ the radial Green's function. The
left-hand side of Eq. (\ref{01}) can then be cast into 
\begin{equation}
\sum_{m=-\infty }^{\infty }\left\{ E+\left[ \frac{\hbar ^{2}}{2\mu }\left( 
\frac{d^{2}}{d\rho ^{2}}+\frac{1}{\rho }\frac{d}{d\rho }-\frac{m^{2}}{\rho
^{2}}\right) \right] -V(\rho )\right\} G_{m}^{0}(\rho ,\rho ^{\prime };E)%
\frac{e^{im(\varphi -\varphi ^{\prime })}}{2\pi }.  \label{002}
\end{equation}
For a charged particle affected by a magnetic field, the Green's function $G(%
{\bf x,x}^{\prime };E)$ is different from $G^{0}({\bf x,x}^{\prime };E)$ by
a global NPF \cite{2,7,8} 
\begin{equation}
G({\bf x},{\bf x}^{\prime };E)=G^{0}({\bf x},{\bf x}^{\prime };E)\exp
\left\{ \frac{ie}{\hbar c}\int_{{\bf x}^{\prime }}^{{\bf x}}{\bf A(\tilde{x}}%
)\cdot d{\bf \tilde{x}}\right\} .  \label{003}
\end{equation}
Here the vector potential ${\bf A(x})$ is used to describe the magnetic
field. For an infinitely thin tube of finite magnetic flux along the $z$%
-direction, the vector potential can be expressed as
\begin{equation}
{\bf A(x})=2g\frac{-y{\hat{e}}_{x}+x{\hat{e}}_{y}}{x^{2}+y^{2}},  \label{05a}
\end{equation}
where ${\hat{e}}_{x},{\hat{e}}_{y}$ stand for the unit vector along the $x,y$
axis respectively. Introducing the azimuthal angle $\varphi ({\bf x})=\tan
^{-1}(y/x)$ around the $AB$ tube, the components of the vector potential can
be expressed as $A_{i}=2g\partial _{i}\varphi ({\bf x}).$ The associated
magnetic field lines are confined to an infinitely thin tube along the $z$%
-axis, 
\begin{equation}
B_{3}=2g\epsilon _{3ij}\partial _{i}\partial _{j}\varphi ({\bf x})=4\pi
g\delta ({\bf x}_{\bot }),  \label{006}
\end{equation}
where ${\bf x}_{\bot }${\bf \ }stands for the transverse vector ${\bf x}%
_{\bot }\equiv (x,y).$ Since the magnetic flux through the tube is defined
by the integral $\Phi =\int d^{2}xB_{3}$, the coupling constant $g$ is
related to the magnetic flux by $g=\Phi /4\pi $. By using the expression of $%
A_{i}=2g\partial _{i}\varphi $, the angular difference between the initial
point ${\bf x}^{\prime }$ and the final point ${\bf x}$ in the exponent of
the NPF is given by 
\begin{equation}
\varphi -\varphi ^{\prime }=\int_{t^{\prime }}^{t}d\tau \dot{\varphi}(\tau
)=\int_{t^{\prime }}^{t}d\tau \frac{-y\dot{x}+x\dot{y}}{x^{2}+y^{2}}=\int_{%
{\bf x}^{\prime }}^{{\bf x}}\frac{{\bf \tilde{x}\times }d{\bf \tilde{x}}}{%
{\bf \tilde{x}}^{2}},  \label{007}
\end{equation}
where $\dot{\varphi}=d\varphi /d\tau $. Given two paths $C_{1}$ and $C_{2}$
connecting ${\bf x}^{\prime }$ and ${\bf x}$, the integral differs by an
integer multiple of $2\pi $. The winding number is thus given by the contour
integral over the closed difference path $C$: 
\begin{equation}
n=\frac{1}{2\pi }\oint_{C}\frac{{\bf \tilde{x}\times }d{\bf \tilde{x}}}{{\bf 
\tilde{x}}^{2}}.  \label{008}
\end{equation}
The magnetic interaction is therefore purely nonlocal and topological \cite
{5,6,7,8,yang}. Its action takes the form ${\cal A}_{{\rm mag}}=-\hbar \mu
_{0}2\pi n,$ where $\mu _{0}\equiv -2eg/\hbar c=-\Phi /\Phi _{0}$ is a
dimensionless number with the customarily minus sign. The NPF now becomes $%
\exp \left\{ -i\mu _{0}(2\pi n+\varphi -\varphi ^{\prime })\right\} $. The
Green's function $G_{n}(\rho ,\rho ^{\prime };E)$ for a specific winding
number $n$ can be obtained by converting the summation over $m$ in Eq. (\ref
{002}) into an integral over $z$ and another summation over $n$ by the
Poisson's summation formula (e.g. Ref. \cite{18} p.469) 
\begin{equation}
\sum_{m=-\infty }^{\infty }f(m)=\int_{-\infty }^{\infty }dz\sum_{n=-\infty
}^{\infty }e^{2\pi nzi}f(z).  \label{06a}
\end{equation}
So the expression (\ref{002}) when includes the NPF can be written as 
\begin{equation}
\int dz\sum_{n=-\infty }^{\infty }\left\{ E+\left[ \frac{\hbar ^{2}}{2\mu }%
\left( \frac{d^{2}}{d\rho ^{2}}+\frac{1}{\rho }\frac{d}{d\rho }-\frac{z^{2}}{%
\rho ^{2}}\right) \right] -V(\rho )\right\} G_{z}(\rho ,\rho ^{\prime };E)%
\frac{e^{i(z-\mu _{0})(\varphi +2n\pi -\varphi ^{\prime })}}{2\pi },
\label{0010}
\end{equation}
where the superscript $0$ in $G_{n}^{0}$ has been suppressed to denote that
the $AB$ effect is included. Obviously, the number $n$ in the right-hand
side is precisely the winding number by which we want to classify the
Green's function. Employing the special case of the Poisson formula $%
\sum_{n}\exp \{ik(\varphi +2n\pi -\varphi ^{\prime })\}=\sum_{m=-\infty
}^{\infty }\delta (k-m)\exp \{im(\varphi -\varphi ^{\prime })\},$ the
summation over all indices $n$ forces $z=\mu _{0}$ modulo an arbitrary
integer number. Thus, we obtain 
\begin{equation}
\sum_{m=-\infty }^{\infty }\left\{ E+\left[ \frac{\hbar ^{2}}{2\mu }\left( 
\frac{d^{2}}{d\rho ^{2}}+\frac{1}{\rho }\frac{d}{d\rho }-\frac{\left| m+\mu
_{0}\right| ^{2}}{\rho ^{2}}\right) \right] -V(\rho )\right\} G_{\left|
m+\mu _{0}\right| }(\rho ,\rho ^{\prime };E)\frac{e^{im\varphi }}{2\pi }.
\label{07}
\end{equation}
We see that the influence of the $AB$ effect to the radial Green's function
is to replace the integer quantum number $m$ with a real one $\left| m+\mu _{0}\right| $ which
depends on the magnitude of magnetic flux. Applying the Fourier expansion of 
$\delta $ function, 
\begin{equation}
\delta (\varphi -\varphi ^{\prime })=\sum_{m=-\infty }^{\infty }\frac{1}{%
2\pi }e^{im(\varphi -\varphi ^{\prime })},  \label{0011}
\end{equation}
to the rhs of Eq. (\ref{01}) and defining $\alpha =\left| m+\mu _{0}\right| $
for convenience, we reduce the radial Green's function to 
\begin{equation}
\left\{ E+\left[ \frac{\hbar ^{2}}{2\mu }\left( \frac{d^{2}}{d\rho ^{2}}+%
\frac{1}{\rho }\frac{d}{d\rho }-\frac{\alpha ^{2}}{\rho ^{2}}\right) \right]
-V(\rho )\right\} G_{\alpha }(\rho ,\rho ^{\prime };E)=\delta (\rho -\rho
^{\prime }).  \label{09}
\end{equation}
As a result, the corresponding radial wave equation reads 
\begin{equation}
\left\{ E+\left[ \frac{\hbar ^{2}}{2\mu }\left( \frac{d^{2}}{d\rho ^{2}}+%
\frac{1}{\rho }\frac{d}{d\rho }-\frac{\alpha ^{2}}{\rho ^{2}}\right) \right]
-V(\rho )\right\} R_{\alpha k}(\rho )=0,  \label{0012}
\end{equation}
where the subscript set $(\alpha ,k)$ with $k\equiv \sqrt{2\mu E}/\hbar $
denotes the state of scattering particle.

For a short range potential, say $V(\rho )$ vanishes as $\rho >a$, the
exterior solution is the linear combination of 1st and 2nd kind Bessel
functions $J_{\alpha }(k\rho )$, and $N_{\alpha }(k\rho )$, and may be given
by 
\begin{equation}
R_{\alpha k}(\rho )=\sqrt{k}\left[ \cos \delta _{\alpha }(k)J_{\alpha
}(k\rho )-\sin \delta _{\alpha }(k)N_{\alpha }(k\rho )\right] ,  \label{010}
\end{equation}
where $k=\sqrt{2\mu E}/\hbar $ and $\delta _{\alpha }(k)$ is the phase shift
which can be used to measure the interaction strength of potential. Thus the
general solution $\Psi _{k}({\bf x})$ of a scattering particle is given by
superposition of the partial waves $\Psi _{\alpha k}({\bf x})=R_{\alpha
k}(\rho )e^{im\varphi }$, which reads 
\begin{equation}
\Psi _{k}({\bf x})=\sum_{m=-\infty }^{\infty }\sqrt{k}\left[ \cos \delta
_{\alpha }(k)J_{\alpha }(k\rho )-\sin \delta _{\alpha }(k)N_{\alpha }(k\rho )%
\right] e^{im\varphi }.  \label{011}
\end{equation}
Since it must describe both the incident and the scattered waves at large
distance, we naturally expect it to become 
\begin{equation}
\Psi _{k}({\bf x})\stackrel{\left| {\bf x}\right| \rightarrow \infty }{%
\longrightarrow }{\cal F}_{\infty }\left( \exp \{i{\bf k\cdot x}\}\exp
\left\{ \frac{ie}{\hbar c}\int_{C}^{{\bf x}}{\bf A(x}^{\prime })\cdot d{\bf x%
}^{\prime }\right\} \right) +f(\varphi )\sqrt{\frac{i}{\rho }}\exp \{ik\rho
\},  \label{012}
\end{equation}
where $\exp \{i{\bf k\cdot x}\}$ describes the incident plane wave of a
charged particle with momentum{\bf \ }${\bf p}=\mu {\bf k}$ and ${\cal F}%
_{\infty }(\cdot )$ stands for its asymptotic form. The phase modulation of
the NPF comes from the fact that the field ${\bf A(x})$ of $AB$ magnetic
flux affects the charged particle globally. The subscript $C$ in the
integral is used to represent the nature of the NPF which depends on the
different paths. To find the amplitude $f(\varphi )$ we first note that the
plane wave in Eq. (\ref{012}) can be expanded in terms of the partial waves 
\begin{equation}
e^{i{\bf k\cdot x}}=\sum_{m=-\infty }^{\infty }i^{m}J_{m}(k\rho
)e^{im\varphi }.  \label{013}
\end{equation}
Using the same procedure as in Eqs. (\ref{06a})-(\ref{07}), we combine the
nonlocal flux effect into the partial wave expansion, and obtain the result 
\begin{equation}
e^{i{\bf k\cdot x}}e^{\frac{ie}{\hbar c}\int_{C}^{{\bf x}}{\bf A(x}^{\prime
})\cdot d{\bf x}^{\prime }}=\sum_{m=-\infty }^{\infty }i^{\alpha }J_{\alpha
}(k\rho )e^{im\varphi }.  \label{014}
\end{equation}
Taking the asymptotic approximations of Bessel functions, and comparing both
asymptotic forms of Eqs. (\ref{011})\ and (\ref{012}), we find the
scattering amplitude 
\begin{equation}
f(\varphi )=\frac{1}{\sqrt{2\pi k}}\sum_{m=-\infty }^{\infty }e^{i(\delta
_{\alpha }-\pi /4)}2i\sin \delta _{\alpha }e^{im\varphi }.  \label{015}
\end{equation}
It is noteworthy that if the flux is quantized for integer $\mu _{0}$, the
result reduces to the flux-free case \cite{lin}. In most cases, the total
cross section of our major concern is defined by 
\begin{equation}
\sigma _{t}=\int_{-\pi }^{\pi }\left| f(\varphi )\right| ^{2}d\varphi .
\label{016}
\end{equation}
Thus, the partial wave representation of total cross section for a charged
particle scattered by a short range potential plus the nonlocal $AB$ effect
is given by 
\begin{equation}
\sigma _{t}=\frac{4}{k}\sum_{m=-\infty }^{\infty }\sin ^{2}\delta _{\alpha }.
\label{017}
\end{equation}
It is obvious that the cross section is completely determined by the
scattering phase shifts which are concluded by the potential of different
types. Furthermore, when a nonlocal $AB$ magnetic flux exists, both the
phase shift and the cross section are affected globally. A relation between
the total cross section $\sigma _{t}$ and the scattering amplitude is
obtained if we set $\varphi =0$, and then take the imaginary part. It gives $%
\sigma _{t}=\left( 2\sqrt{2\pi }/\sqrt{k}\right) 
\mathop{\rm Im}%
f(0)$. This is the optical theorem and is essentially a consequence of the
conservation of particles. For the case of identical bosons (fermions)
carrying the magnetic flux in two dimensions, the differential cross section
is given by $\sigma (\varphi )=\left| f(\varphi )\pm f(\varphi +\pi )\right|
^{2}$, where the plus sign is for bosons as usual. The total cross sections
are given by the integral $\int_{-\pi }^{\pi }\sigma (\varphi )d\varphi $,
and yield 
\begin{equation}
\sigma _{t}({\rm bosons})=\frac{16}{k}\sum_{m=-\infty ,{\rm even}}^{\infty
}\sin ^{2}\delta _{\alpha },  \label{018}
\end{equation}
and 
\begin{equation}
\sigma _{t}({\rm fermions})=\frac{16}{k}\sum_{m=-\infty ,{\rm odd}}^{\infty
}\sin ^{2}\delta _{\alpha },  \label{019}
\end{equation}
where the subscript ``odd'' (``even'') is used to indicate the summation
over odd (even) numbers only. As a realization of the nonlocal influence of
the $AB$ flux on the cross section, let us consider a charged particle
scattered by a hard disk potential and a magnetic flux. The potential is
given by $V(\rho )=\infty ,$ for $\rho \leq a$, and $V(\rho )=0,$ for $\rho
\leq a$. Using the boundary condition of the wave function $R_{\alpha
k}(a^{+})=0$, we find that the phase shift is given by 
\begin{equation}
\tan \delta _{\alpha }(k)=\frac{J_{\alpha }(ka)}{N_{\alpha }(ka)},
\label{020}
\end{equation}
where $J_{\alpha }(z)$ ($N_{\alpha }(z))$ is the Bessel function of first
(second) kind. Substituting this expression into Eq. (\ref{017}), the total
cross-section is found to be 
\begin{equation}
\sigma _{t}=\frac{4}{k}\sum_{m=-\infty }^{\infty }\frac{J_{\alpha }^{2}(ka)}{%
J_{\alpha }^{2}(ka)+N_{\alpha }^{2}(ka)}.  \label{022}
\end{equation}
Note that the result will reduce to the pure disk case if the flux is
quantized for $\mu _{0}=n\Phi _{0}$. In this case the low energy limit $%
k\rightarrow 0$ (assuming the radius $a$ is finite) can be found by the
asymptotic expansion of Bessel functions, and only index $m=0$ survives. So
the phase shift becomes 
\begin{equation}
\tan \delta _{0}(k)=J_{0}(ka)/N_{0}(ka)\approx \frac{\pi }{2\ln (ka/2)}.
\label{022a}
\end{equation}
This implies the total cross section at the low energy limit is 
\begin{equation}
\sigma _{t}\approx \frac{8a}{\pi }\frac{1}{ka\ln (ka)}\longrightarrow \infty 
\text{.}  \label{023}
\end{equation}
At the high energy limit $k\rightarrow \infty $, we may use the formulas of
Bessel functions of the large argument to turn Eq. (\ref{022}) into 
\[
\sigma _{t}=\frac{4}{k}\sum_{m=-\infty }^{\infty }\cos ^{2}\left[ ka-\left(
m+\frac{1}{2}\right) \frac{\pi }{2}\right] 
\]
\begin{equation}
=\lim_{ka\rightarrow \infty }\frac{4}{k}\left\{ \sum_{m=-[ka],{\rm even}%
}^{[ka]}\cos ^{2}\left( ka-\frac{\pi }{4}\right) +\sum_{m=-[ka],{\rm odd}%
}^{[ka]}\sin ^{2}\left( ka-\frac{\pi }{4}\right) \right\} =4a.  \label{024}
\end{equation}
The value $4a$ explains that the quantum result does not go over to the
actual classical result $\sigma _{t}\rightarrow 2a$ even though the wave
length of de Broglie is much less than $a$. The numerical result for $\alpha 
$ with noninteger value is plotted in Fig. 1, where the normalization $%
\sigma _{0}$ is chosen as $4a$. There are two main results: (1) The cross
section $\sigma _{t}$ is drastically suppressed at the low energy limit
(equivalently, the short range potential), say $ka\leq 1$, at quantized
magnetic flux $\Phi =(2n+1)\Phi _{0}/2$, $n=0,1,2,\cdots $, with $\Phi _{0}$
periodicity as shown in Fig. 1 and Fig. 2. (2) A more interesting
consideration is given by the scattering of identical particles simulated by
the hard disks carrying the magnetic flux. In Fig. 3, we plot the total
cross sections of identical bosons carrying the magnetic flux via Eq. (\ref
{018}). The outcome shows that the cross section approaches zero ($\sigma
_{t}\rightarrow 0$) when the value $ka\rightarrow 0$ if the magnetic flux is
at quantized value $(2n+1)\Phi _{0}$. On the contrary, if the magnetic flux
is equal to $2n\Phi _{0}$, the cross section becomes maximum and the effect
of magnetic flux disappears. Since the decay rate of a current ${\bf j}$
traveling a distance ${\bf x}$ is given by ${\bf j(x)=j(}0{\bf )}\exp
(-\sigma _{t}n_{0}{\bf x})$, where $n_{0}$ is the number of the scattering
center, the total cross section $\sigma _{t}\rightarrow 0$ at the low energy
limit at $\Phi =(2n+1)\Phi _{0}$ means that the resistance $R\rightarrow 0$
and results in the persistence of current. This phenomenon is consistent
with the picture of composite boson in fractional quantum Hall states
located at the filling factor with odd denominator such as $\nu =1/3$. The
composite boson is pictured by an electron carrying the quantized magnetic
flux $\Phi =(2n+1)\Phi _{0}$. It dictates the quantized Hall states which
exhibit the perfect conduction in the longitudinal direction, i.e. the
resistance originated from the collisions between composite bosons disappear 
\cite{10}. The global structure of the total cross section is given by $%
2\Phi _{0}$ periodicity as shown in Fig. 4. In the case of identical
fermions, the total cross section $\sigma _{t}\rightarrow 0$ is found at the
quantized magnetic flux $\Phi =2n\Phi _{0}$ as shown in Fig. 5. Such effect
is consistent with the model of composite fermion in the quantum Hall state
located at the filling factor with even denominator $\nu =5/2$. The
composite fermion is described by an electron carrying the quantized
magnetic flux $\Phi =2n\Phi _{0}$. In Ref. \cite{20}, a quantitative
explanation of quantum Hall state at the filling factor $\nu =5/2$ is given
by the existence of a shorter range potential between the composite fermions
than the case of the filling factor $\nu =1/2$. Here we can see that, in
Fig. 5, a sufficiently short range potential, say $ka<0.5$, between the
fermions carrying the quantized magnetic flux $\Phi =2n\Phi _{0}$ will cause
negligible cross section and thus agree with the composite fermions model.
Similar to the boson case, the oscillating period is given by $2\Phi _{0}$
as shown in Fig. 6.

In this paper, we study the partial wave method of scattering theory for a
short range potential and a magnetic flux. As an illustration, the hard disk
potential plus a magnetic flux is calculated in detail. The nonlocal
influence of the magnetic flux is discussed. Since the nonlocal effect of
magnetic flux on the charged particle is universal, the effect should be
general in similar systems. Although we assume that the potential must be $%
V(\rho )=0$ for $\rho >a$, we do not specify the radius $a$ beyond which $%
V(\rho )=0$. Hence we expect that the method given in this paper should be
valid for a very general potential as long as the potential decreases
rapidly enough when $r\rightarrow \infty $. On the other hand, though in our
discussion the magnetic flux is placed at the origin, it can be moved to the
other points as long as it still locates in the potential region. This is
due to the fact that the final outcome just relates to the flux via homotopy
classes. We hope the discussions would be helpful in understanding
mesoscopic systems and strongly correlated systems.

\centerline{ACKNOWLEDGMENTS} 
\center{The author would like to thank the referee for his comments,
professor Pi-Guan Luan for helpful discussions, Professor Jang-Yu Hsu, and Dr. Y.N. Chen for reading the manuscript.}
\newpage
\begin{figure}[hbt]\includegraphics[width=2.8in]{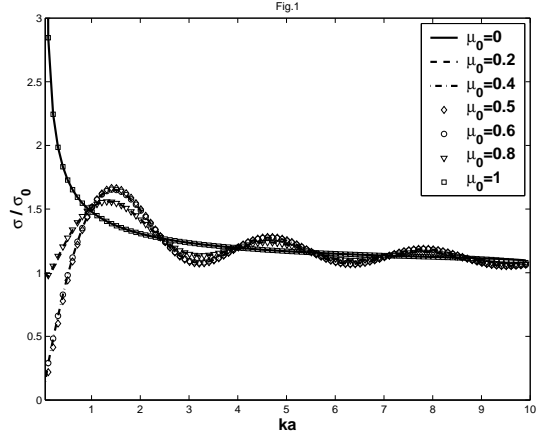}
\caption{ The total cross section for a charged particle scattered by a hard
disk with radius $a$ and a magnetic flux along the z-axis. The
normalization $\protect\sigma_0=4a$ has been selected. Due to the existence
of magnetic flux, at the limit of the long wave
(equivalently, the short range potential), say $ka \leq 1$, the total corss section is drastic suppressed
at quantized magnetic flux $\Phi =(2n+1)\Phi _{0}/2$, where $n=0,1,2,\cdots $, with $\Phi _{0}$ periodicity, see Fig. 2.
The magnetic flux effect disappears when
the flux is quantized at $\Phi=n\Phi _{0}$.}
\end{figure}

\begin{figure}[hbt]\includegraphics[width=2.8in]{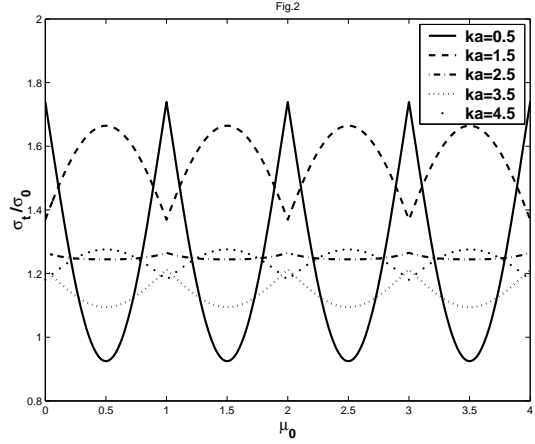}
\caption{Periodic structures of total cross sections of
a charged particle scattered by
a hard disk plus a magnetic flux along the z-axis.
At quantized values of magnetic flux $\Phi=(2n+1)\Phi _{0}/2$, $n=0,1,2,\cdots $, the
cross section reduces to the minimum for $ka \leq 0.5$.}
\end{figure}

\begin{figure}[hbt]\includegraphics[width=2.8in]{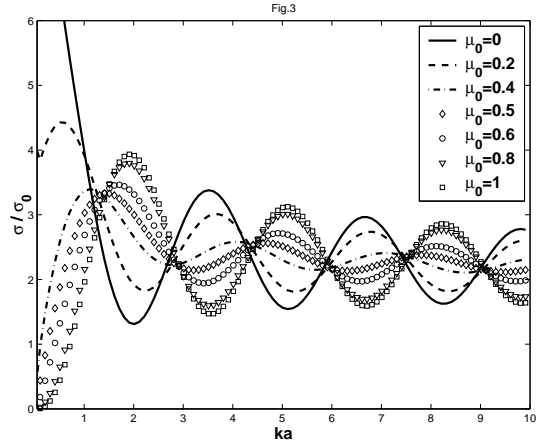}
\caption{Total cross sections for identical bosons carrying the
magnetic flux with various $\protect\mu_0$. The cross section at the
long wave length limit (equivalently, the sufficient short range potential), say $ka \leq 0.5$,
approaches zero at the quantized magnetic flux
$\Phi=(2n+1)\Phi _{0}$.
On the contrary, the cross section becomes maximum and the effect of magnetic flux disappears
when $\Phi=2n\Phi _{0}$.
The periodic structure is $2\Phi _{0}$ as shown in Fig. 4.}
\end{figure}

\begin{figure}[hbt]\includegraphics[width=2.8in]{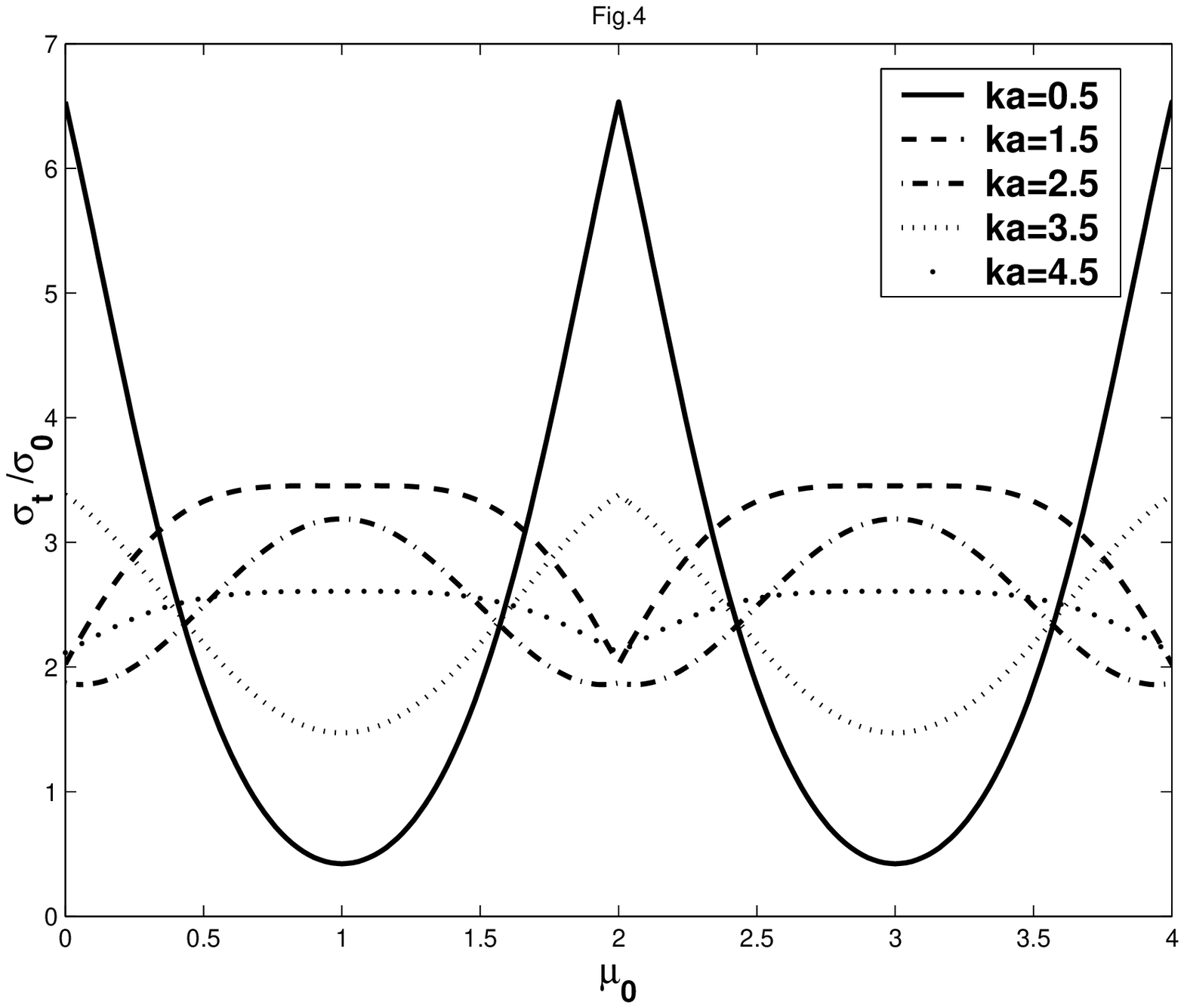}
\caption{Periodic structures of cross sections of
identical bosons carrying the magnetic flux. The cross section approaches zero
when the magnetic flux is quantized at $\Phi=(2n+1)\Phi_{0}$ for $ka \leq 0.5$.}
\end{figure}

\begin{figure}[hbt]\includegraphics[width=2.8in]{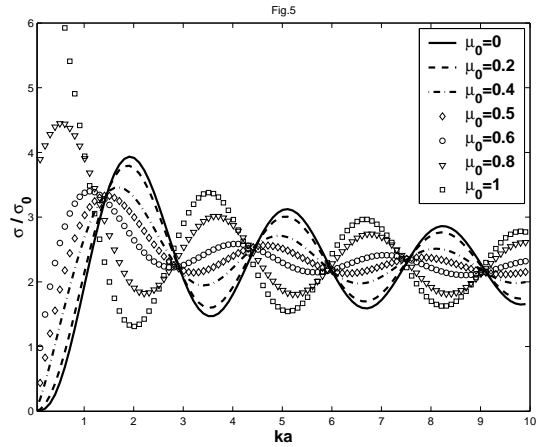}
\caption{Total cross sections of identical fermions carrying the
magnetic flux with various $\protect\mu_0$. The
cross section approaches zero for $ka \leq 0.5$ when the flux becomes
$2n\Phi _{0}$. The magnetic flux effect disappears
when the magnitude of flux is at $(2n+1)\Phi _{0}$.
The global periodic structures in cross sections is $2\Phi _{0}$ as shown in Fig. 6.}
\end{figure}

\begin{figure}[hbt]\includegraphics[width=2.8in]{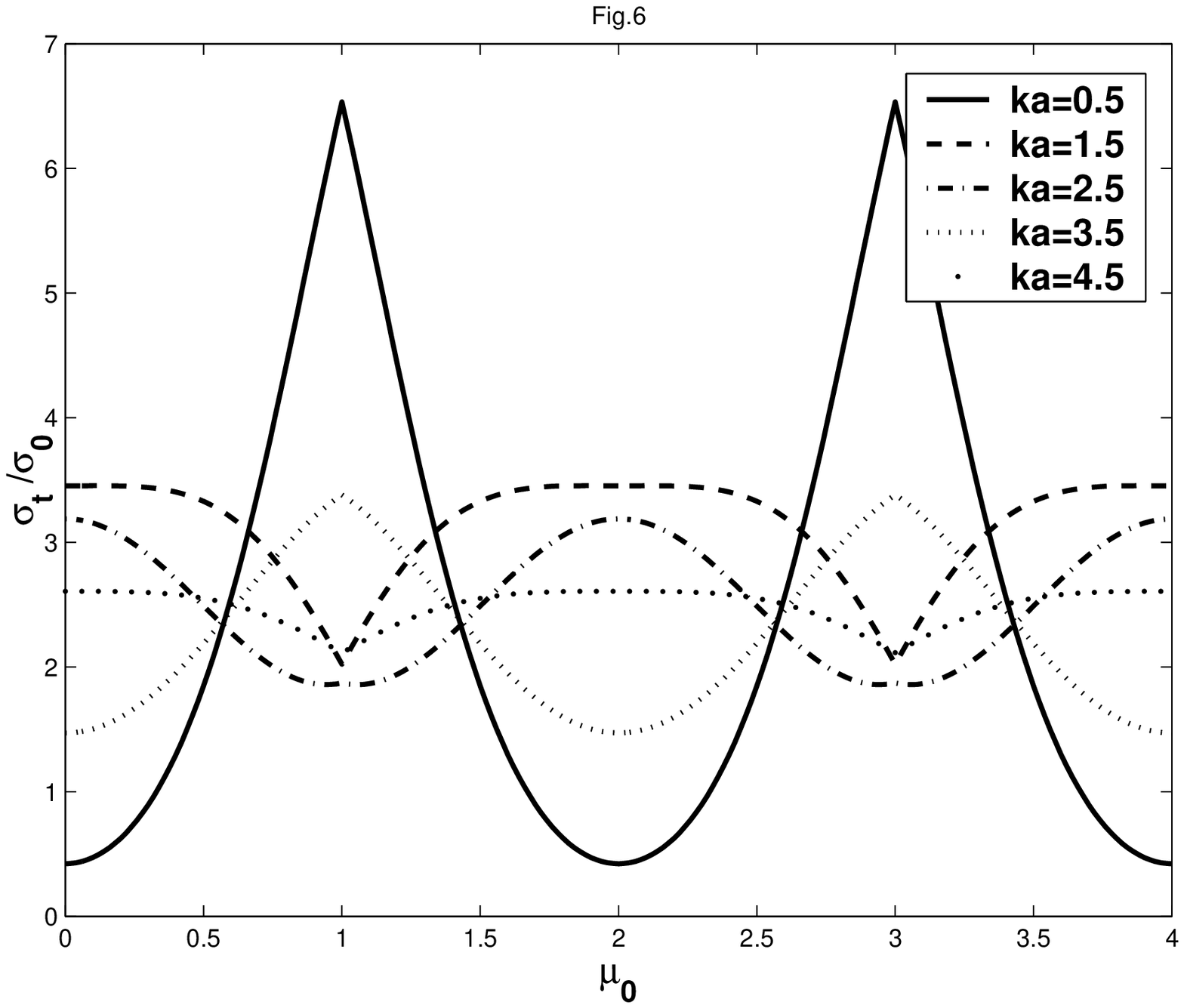}
\caption{Periodic structures of total cross sections
for identical fermions carrying the magnetic flux. The cross section approaches
zero when the magnetic flux is quantized at $\Phi=2n\Phi _{0}$ for $ka \leq 0.5$.}
\end{figure}
\end{document}